\documentclass[twocolumn,superscriptaddress,showpacs,amsmath,amssymb,aps,prl,floatfix]{revtex4}

\usepackage{graphicx}
\usepackage{dcolumn}
\usepackage{bm}

\usepackage[usenames,dvipsnames]{xcolor}
\usepackage{dsfont}
\usepackage{amsbsy}

\usepackage{amsfonts}
\usepackage{graphicx}
\usepackage{bbold}

\usepackage{calrsfs}
\DeclareMathAlphabet{\pazocal}{OMS}{zplm}{m}{n}

\makeatletter
\def\hlinewd#1{%
  \noalign{\ifnum0=`}\fi\hrule \@height #1 \futurelet
   \reserved@a\@xhline}
\makeatother

\begin{document}
\title{Refocusing  dipolar interactions between electronic spins of donors in silicon}

\author{T.~S.~Monteiro}
\affiliation{Department of Physics and Astronomy, University College London,
Gower Street, London WC1E 6BT, United Kingdom}

\date{\today}

\begin{abstract}

We note the existence of a set of magnetic field values where a simple Hahn echo sequence refocuses the dynamics of the full dipolar interaction, for spin systems of  electron donors in silicon.  As the refocussing occurs for both arbitrary coupling strengths and arbitrary pulse timings, these dipolar refocusing points (DRPs) offer new possibilities for regulating entanglement due to the always-on spin dipolar interaction. While the experimental effects of DRPs will be strongly diluted in the measured coherences of thermal (unpolarized) spin ensembles,  we investigate possible signatures in coherence decays arising from a study of the combined effects of decoherence arising from 
instantaneous diffusion and direct flip-flops.
\end{abstract}




\pacs{03.65.Yz, 03.67.Lx}

\maketitle

\begin{figure}[htb]
\includegraphics[width=2.3in]{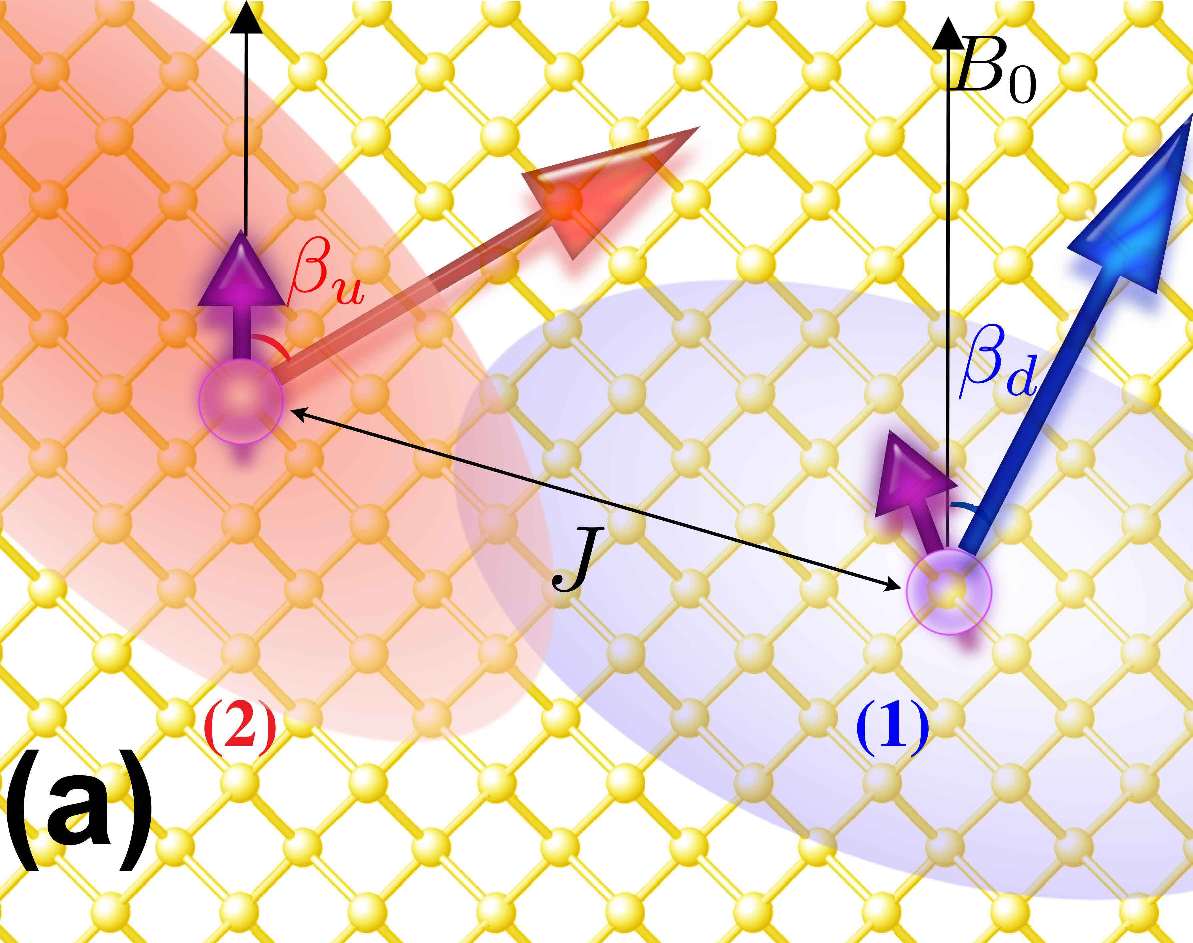}
\includegraphics[width=3.in]{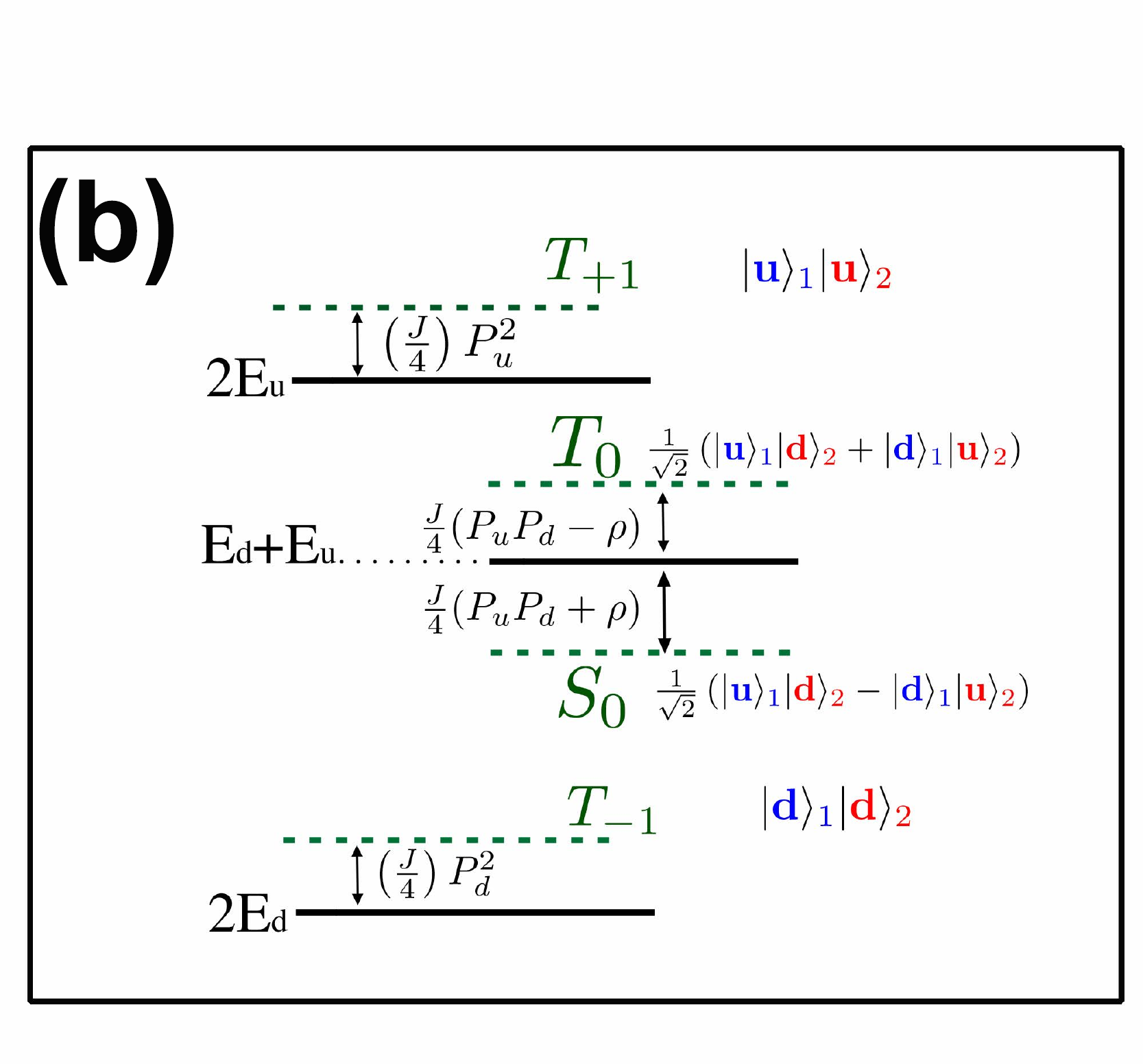}
\caption{ {\bf (a)} Illustrates interactions between two donor spin systems. The electronic spins $s=1/2$, but because of strong coupling to the host nuclear spins (purple arrows) the  ${\hat S}_z$ eigenvalues $m_s=\pm 1/2$ are not good quantum numbers. Instead, each spin quantum state $i$ corresponds to eigenstates of an effective field, tilted to the $z$-axis by an angle $\beta_i$, with $P_i= \cos \beta_i= 2\langle i |{\hat S}_z|i\rangle$. Microwave pulses resonantly drive transitions $i= u \to d$
between two selected states; the figure illustrates two donor atoms in the spin state 
$|\psi\rangle=|d\rangle_1 |u\rangle_2$. At values of $B_0$ corresponding to {\em dipolar refocusing points} (DRPs), 
a Hahn echo sequence completely refocuses the dynamical effect of the full dipolar coupling
${\hat H}_D= J {\hat S}_{z1}{\hat S}_{z2} -\frac{J}{4}({\hat S}^+_{1}{\hat S}^-_{2}+{\hat S}^-_{1}{\hat S}^+_{2})$
on the $|u\rangle_1 |d\rangle_2$ and  $|d\rangle_1 |u \rangle_2$ states, e.g.
$|u\rangle_1 |d\rangle_2 \to [ (\pi/2)_y - \tau -\pi - \tau-(\pi/2)_y]  \to |u\rangle_1 |d\rangle_2$,
for arbitrary $J$ and $\tau$.
 {\bf (b) } Represents the energy levels of an interacting pair of spin systems. 
At high magnetic fields, without mixing, the states of the $u \to d$ transition yield a triplet of states ($T_{+1},T_0, T_{-1}$) and a singlet $S_0$. With mixing,
a similar structure is preserved, but the energy shifts due to $H_D$ (shown in the figure) are 
field-dependent and are given by 
$P_u,P_d$ and $\rho=\langle u d |{\hat S}^+_{1}{\hat S}^-_{2}+{\hat S}^-_{1}{\hat S}^+_{2}| d u \rangle$.
DRPs occur at field values where $P^2_u+P^2_d=2(P_uP_d+\rho)$. }
\label{Fig1}
\end{figure}

Spin qubit systems can interact and even entangle with other neighbouring spins through
 their magnetic dipolar coupling. More broadly,
this interaction is of key importance in NMR (nuclear magnetic resonance)
 and ESR (electron spin resonance) studies \cite{Schweiger}. 
 After Hahn's seminal discovery that the spin echo sequence \cite{Hahn} 
can eliminate and refocus the effects of static magnetic field inhomogeneities,
an important early study showed that more complex, rapid 
``dynamical decoupling'' sequences of pulses with alternating orientations
 could approximately reverse the effects of dipolar coupling \cite{Rhim1970}.
Nevertheless, controlling or regulating this ``always-on'' interaction 
remains a challenge in a range of
potential architectures in the field of quantum information \cite{Nielsen}. 

The last few years have seen rapid advances in the  coherent manipulation of single or few spin systems, 
both in diamond NV colour centers \cite{NV1,NV2,NV2a,NV3,NV4,NV5} as well as with phosphorus donors in silicon \cite{Morello2010,Pla2012,Pla2014}.  By careful control of pulse timing, dipolar 
coupling of known strength has been used to generate maximally
entangled states \cite{Neumann} of NV colour centres. In the case of donors, most studies are in
  on ensembles where exceptionally long coherence times have been demonstrated
 in isotopically pure samples \cite{Tyryshkin2012,Steger2012}. For such pure samples, 
decoherence is dominated by interactions 
between dipolar coupled donor spins, rather than being dominated by the 
flip-flopping of $^{29}$Si nuclear spin impurities in natural silicon which leads to dephasing
decoherence ($T_2$ times).

There is also growing interest in other donor species such as bismuth, arsenic and antimony where there is
strong mixing between the donor electronic and nuclear spins \cite{George2010,Morley2010}.
For these, Optimal Working Points (OWPs)  of enhanced electronic coherence have been identified, even in natural silicon, first investigated theoretically
\cite{Mohammady2010,Mohammady2012,Balian2012}, then subsequently including experiments \cite{Wolfowicz2013,Balian2014}.
At OWPs, there is suppression of dephasing processes (arising from diagonal interactions between spins such as 
${\hat S}_{z1}{\hat S}_{z2}$ terms).

However, we show here that there is also a quite distinct set of field values 
where the full dipolar coupling can be eliminated, by a different mechanism involving the off-diagonal
interactions, for arbitrary times and coupling strengths, without the need for any complex dynamical decoupling sequences.
 A simple echo pulse suffices, since at these ``magic'' field values, which we refer to below as  dipolar refocusing points (DRPs), 
the donor spins' own internal level structure can lead to complete destructive interference
 between diagonal and off-diagonal dipolar contributions, with no requirement for pulse timing to be fast
compared with the internal dynamics.
While OWPs  have been observed in other physical systems such as in flux qubits or clock transitions of atoms, to our
knowledge, DRPs do not. The effects of DRPs are not however as readily visible with thermal ensembles as the OWPs.
Thus the most practical applications may only become realisable when single spin or few spin techniques tested on phosphorus are extended to other donor species in silicon or if there are alternative advances which permit coherent control of donors at the same level
as NV centres. Nevertheless, we discuss here the possibility of signatures in coherence decays in isotopically  pure samples of$^{28}$Si.

We consider a spin system with eigenstates which are given by the donor spin Hamiltonian:
\begin{equation}
\hat{H}_0 |i\rangle = E_i  |i\rangle.
\label{Eq:donor1}
\end{equation}
Like many other types of spin qubit systems,
the donor spins interact primarily via the dipolar interaction between their electronic spins,
which in its secular form is given by:
\begin{equation}
{\hat H}_D= J \left [{\hat S}_{z1}{\hat S}_{z2} -\frac{1}{4}({\hat S}^+_{1}{\hat S}^-_{2}+{\hat S}^-_{1}{\hat S}^+_{2})\right]= H_{zz} +H_{ff}
\label{Eq:dipolar}
\end{equation}
which represents the sum of a diagonal (Ising) term and a  spin flip term.
We consider qubits restricted to a  two-state space $i=u,d$ resonantly coupled by microwave radiation of frequency  $\hbar f_{u \to d}=E_u-E_d$.
Under the action of the Hamiltonians $\hat{H}_0+{\hat H}_D$, we can define a four-state space,
$|T_{+1}\rangle=|u\rangle_1 |u\rangle_2$,  $|T_{-1}\rangle=|d\rangle_1 |d\rangle_2$, plus
$|T_{0}\rangle=\frac{1}{2} [|u\rangle_1 |d\rangle_2 + |d\rangle_1 |u \rangle_2]$ and
$|S_{0}\rangle=\frac{1}{2} [|u\rangle_1 |d\rangle_2 - |d\rangle_1 |u \rangle_2]$.
The four eigenstates are labelled using the triplet/singlet classification of spin-$1/2$ pair states. However,
we are considering here  spin systems  which have electron spin $S=1/2$ but also strong coupling with the host nucleus and a  nuclear spin $I  > 1/2$. These include bismuth, arsenic and antimony  but
 not the very well-studied phosphorus qubit Si:P system which does not have these essential attributes. 
For modest magnetic fields $B_0$,  the Zeeman quantum states $|m_s,m_I \rangle$ are not eigenstates of ${\hat{H}}_0$.
Instead,  $\langle i |{\hat S}_z|i\rangle = \frac{P_i} {2}$ can take any value $\in [-1/2:1/2]$ 
 and can vary rapidly with magnetic field $B_0$. At high $B_0$, though, $m_s=\pm 1/2$ are good quantum numbers and the triplet/singlet states tend to their canonical spin-$1/2$ forms.

The resulting spectrum is illustrated in Fig.\ref{Fig1}. We can show (see below) that
$\langle uu| {\hat H}_{ff}| uu\rangle = \langle dd| {\hat H}_{ff}| dd\rangle=0$ while
$\langle ud| {\hat H}_{ff}| du\rangle =\langle du | {\hat H}_{ff}| ud\rangle=\frac{J}{4} \rho$
 (analytical forms of $\rho$ as a function of $B_0$ for a given state $i=u,d$ are also given below).
Conversely, $\langle ud | {\hat H}_{zz}| du\rangle =\langle du | {\hat H}_{zz}| ud\rangle=0$ while
$\langle i j| {\hat H}_{zz}| i j\rangle=\frac{J}{4} P_i P_j  $ for any $i,j \equiv u,d$.
From the above, we can easily write down the time-evolved form of the eigenstates:
\begin{eqnarray}
|T_{+1}(t=0) \rangle \to |T_{+1}(t) \rangle= e^{-i(2E_u+ (J/4)P_u^2)t} |u\rangle_1 |u\rangle_2\nonumber \\
|T_{-1 }(t=0) \rangle \to |T_{-1}(t) \rangle= e^{-i(2E_d+ (J/4)P_d^2)t}  | d\rangle_1 |d\rangle_2
\end{eqnarray}
thus the $\pm 1$ triplet states do not lead to entanglement. 

However, if the qubits are
prepared in either the $|u\rangle_1 |d\rangle_2 $ or $ |d\rangle_1 |u \rangle_2$ state, i.e
$|T_{0}\rangle \pm |S_{0}\rangle$ superpositions, the qubits become entangled, e.g.:
\begin{eqnarray}
|u\rangle_1 |d\rangle_2 \to \cos{(J\rho t)} \ |u\rangle_1 |d\rangle_2 +i \sin{(J \rho t)} \ |d\rangle_1 |u\rangle_2
\label{Eq:tangle}
\end{eqnarray}
and likewise for  $|d\rangle_1 |u\rangle_2$. It is this evolution which is eliminated by an echo sequence
at dipole refocusing points (DRPs) i.e. particular field values $B_0\equiv B_{DRP}$;
  more generally, at DRPs, the states 
$T_{+1}, T_{-1}, S_{0}$ become a decoherence-free subspace. 

This is easily seen by considering the
effect of the basic Hahn sequence $(\pi/2)_y - \tau -(\pi)_{x/y} - \tau-(\pi/2)_y$ on either
$ |u\rangle_1 |d\rangle_2$ or $ |d\rangle_1 |u\rangle_2$. The effect of the first $(\pi/2)_y$ pulse on
the former  is 
 $ |u\rangle_1 |d\rangle_2 \to  \frac{1}{2}(|u\rangle_1+ |d\rangle_1) (|u\rangle_2- |d\rangle_2)= 
\frac{1}{2}(|T_{+1}\rangle  - |T_{-1}\rangle+ \sqrt{2} |S_{0}\rangle)$ at $t=0$. This state evolves in time as follows:

\begin{eqnarray}
&2\psi(t)&= e^{-i(2E_u+ \frac{J}{4}P_u^2)t} |u\rangle_1 |u\rangle_2 +
         e^{-i(2E_d+ \frac{J}{4}P_d^2)t}  | d\rangle_1 |d\rangle_2 \nonumber \\
       & + & e^{-i[(E_d+E_u)+ \frac{J}{4}(P_d P_u+\rho)]t} ( | u\rangle_1 |d\rangle_2-|d\rangle_1 |u\rangle_2)
\end{eqnarray}
Then, the $\pi$ pulse and subsequent evolution results in the state, at $t=2\tau$ (but for 
any $\tau$):

\begin{eqnarray}
&2\psi(2\tau)&= e^{-i \frac{J}{4}(P_u^2+P_d^2)\tau}  (|u\rangle_1 |u\rangle_2 - | d\rangle_1 |d\rangle_2) \nonumber  \\ 
                  &-& e^{-i \frac{J}{2}(P_d P_u+\rho)\tau}  (|u\rangle_1 |d\rangle_2 - | d\rangle_1 |u\rangle_2 
\end{eqnarray}

where we disregard the inconsequential global phase $e^{-2i(E_u+E_d)\tau} $. We see that if 
$P^2_u+P^2_d=2(P_uP_d+\rho)$, we obtain
$\psi(2\tau)= \frac{1}{2}(|u\rangle_1- |d\rangle_1) (|u\rangle_2+ |d\rangle_2)$.
Then, the final $(\pi/2)_y$ pulse completely restores the initial state $ |u\rangle_1 |d\rangle_2$.
Thus the entanglement of an initial state $ |u\rangle_1 |d\rangle_2$ or $ |d\rangle_1 |u\rangle_2$
can be fully controlled: if one (or repeated) Hahn pulses are applied, the effect of the dipolar interaction is eliminated.
If they are not, entanglement occurs as in Eq.\ref{Eq:tangle}.
We see also that since the $T_{+1}, T_{-1}$  states each acquire a phase $\phi_1=\frac{J}{4}(P_u^2+P_d^2)$
and the $S_0$ state acquires an identical phase $\phi_0= \frac{J}{2}(P_d P_u+\rho)=\phi_1$,
any arbitrary superposition of $T_{+1}, T_{-1}, S_{0}$ never dephases, regardless of $J$ or $\tau$.

The case for the state $T_{0}$ is different; this state has  phase 
$ \frac{J}{2}(P_d P_u-\rho)$ hence acquires a phase difference $=J\rho$ relative to the other three states.

We now calculate the values of $B_{DRP}$ for each transition.
The Hamiltonian of the donor spin systems Eq.\ref{Eq:donor1} takes the form:
\begin{equation}
\hat{H}_0= \omega_0\gamma_e \left( \hat{S}_z   - \delta  \hat{I}_z \right) +
A{\bf I} \cdot {\bf S},
\label{Eq:donor}
\end{equation}
where $\omega_0=B_0\gamma_e$ and $\gamma_e$ is the electronic gyromagnetic ratio, 
$\delta$ is the ratio of nuclear and electronic gyromagnetic ratios
and $A$ is the  isotropic hyperfine coupling.   For example, for 
Si:Bi,  $\frac{A}{2\pi}=1475.4 $~MHz while the $^{209}$Bi nuclear spin $I=9/2$
and $\delta = 2.488 \times 10^{-4}$.
In this case for magnetic fields $B \lesssim 0.3$T, we have the strong mixing regime
where $B_0\gamma_e= \omega_0 \sim A$. 
Then, the Zeeman quantum states $|m_s,m_I \rangle$ are not eigenstates of $\hat{H}_0$, but
the total $m=m_s+m_I$ is a good quantum number. 
We  consider magnetic fields where $A\gg \omega_I$, with  $\delta \ll 1$  and where the internuclear dipolar coupling is negligible thus we do not consider nuclear spin flips. We note also that the regimes we investigate are distinct from a recent studies of  phosphorus dimers 
which are close enough ($\sim6$ nm) to each other to be exchange-coupled and also have with $J \gg A$
\cite{dimers}; here we consider only dipolar coupled donors with $J \ll A$.

The eigenstates and eigenvalues of donors in such regimes were previously investigated \cite{Mohammady2010, Mohammady2012}.
 There are $(2I+1)(2s+1) $ eigenstates,
ranging from 8 in total for arsenic to 20 states for bismuth for example. There are always two states for
which $|m|=s+I$ are aligned along the z-axis and remain unmixed. The other eigenstates  form doublets of constant $m$:
\begin{eqnarray}
&|+,m \rangle &=  \  \ \cos{\frac{\beta_m}{2}} |\frac{1}{2}, m-\frac{1}{2}\rangle + \sin{\frac{\beta_m}{2}} |-\frac{1}{2}, m+\frac{1}{2}\rangle \nonumber \\
&|-,m \rangle &=  -\sin{\frac{\beta_m}{2}} |\frac{1}{2},m-\frac{1}{2}\rangle+ \cos{\frac{\beta_m}{2}} |-\frac{1}{2}, m+\frac{1}{2}\rangle \nonumber\\
\label{mixed}
\end{eqnarray}
i.e, transformation from the Zeeman basis $|m_s, m_I\rangle$ to/from the eigenstate basis 
$|\pm,m \rangle$ are given by the rotation matrices  $R^T_y(\beta_m)$ and $R_y(\beta_m)$. 
Defining parameters $X_m=I(I+1)-m^2+1/4$ and $Z_m=m+\frac{\omega_0}{A}(1+\delta)$, then
the rotation angles  $\beta_m$ are given analytically by $\beta_m= \tan^{-1}X_m/Z_m$.

\begin{figure}[htb]
\includegraphics[width=3.3in]{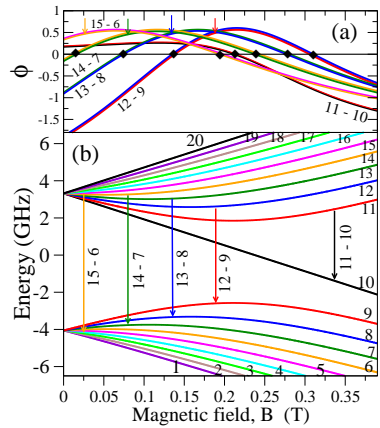}
\caption{ {\bf (a)} Position of DRPs for a few key ESR lines of bismuth.
 Figure plots  function $\phi$ in Eq.\ref{DRP} and crossings of the $x$-axis (black diamonds) denote the DRPs. The two (or one) DRPs (which permit full dipolar
interaction suppression) for each transition
are at different field values to OWPs where only diagonal, dephasing interactions are suppressed. The positions of
the OWP for each transition are shown by arrows; the $14 \to 7$ OWP has been studied experimentally in \cite{Wolfowicz2013}. {\bf (b) } Shows the energy level spectrum for bismuth. The transitions illustrated in (a) are indicated by arrows (at the corresponding OWP position, except for the 11-10 line which has a DRP but no OWP). Doublets of constant $m$ are plotted in the same colour. }
\label{Fig2}
\end{figure}

It can be seen from Eq.\ref{mixed} that
 $\langle \pm,m |{\hat S}_z| \pm,m\rangle= \frac{1}{2} \cos{\beta_m}=\frac{P_m}{2}$.
ESR transitions obey the selection rules $m_s-m'_s=\pm1, m_I-m'_I=0$ thus only  components of the 
$|\pm,m \rangle$ states which obey those same selection rules contribute to the line strength.
For transitions between states $u=|+,m\rangle \to d=|-,m-1\rangle$ 
which are dipole-allowed  at all fields,
we obtain:
\begin{equation}
\rho=\langle u d |{\hat S}^+_{1}{\hat S}^-_{2}+{\hat S}^-_{1}{\hat S}^+_{2}| d u \rangle= 
  \cos^2{\frac{\beta_u}{2}} \cos^2{\frac{\beta_d}{2}}
\end{equation}

Thus the DRP condition $P^2_u+P^2_d-2(P_uP_d+\rho)=(P_u-P_l)^2-2\rho=0$ becomes:

\begin{equation}
 \phi= (\cos{\beta_u}-\cos{\beta_d})^2- 2\cos^2{\frac{\beta_u}{2}} \cos^2{\frac{\beta_d}{2}}=0
\label{DRP}
\end{equation}
and to find the DRP points, we can search for solutions of the above as a function of magnetic field $\omega_0$.
In \cite{Mohammady2012} two types of forbidden transitions which have appreciable transition strengths at 
$B_0 \lesssim 0.3$T were identified; one class turns into NMR transitions at high-fields, the other is completely
forbidden as $B_0 \to \infty$. For the former, $\rho= \cos^2{\frac{\beta_u}{2}} \sin^2{\frac{\beta_d}{2}}$
while for the latter, $\rho= \sin^2{\frac{\beta_u}{2}} \sin^2{\frac{\beta_d}{2}}$; in these cases, the
form of Eq.\label{DRP} must be adjusted, accordingly, to find the corresponding DRPs.

In Fig.\ref{Fig2} we illustrate both the energy level spectrum and the DRPs for some of the key
(i.e. ESR-allowed at all fields) transitions of bismuth; some of these transitions have an Optimal Working 
Point where dephasing effects are suppressed; in particular the $14 \to 7$ ESR  transition has been investigated experimentally (including simulations)
 in \cite{Wolfowicz2013,Balian2014}. 
We note that while these transitions have a single OWP, they have two DRPs.
The $11 \to 10$ transition is of especial interest. The $i=10$ state is one of the unmixed states 
and of especial significance since efficient optical polarisation to that state has been demonstrated 
\cite{Thewalt2010}. This transition has no OWP, but has a DRP at $B_0=0.21$T. At this point,
for state $11$, the   angle $\beta=\pi/2$ (hence $\langle S_z \rangle=0$) while
 for state $10$ then $\beta=0$.

 
The DRP condition however is never satisfied for a pair of NV spins. Nor is it satisfied for Si:P which has
 neither OWPs (associated with its ESR
transitions) or DRPs.
And while there have been ground-breaking experimental studies at the single-spin level for
both of the above, for donors like Si:Bi and Si:As, only studies with ensembles (most thermal
ensembles) are available. Thus we discuss now whether some indirect signatures of DRPs
may be observable from the coherence decays of these systems. For natural silicon, spectral diffusion
arising from flip-flopping of nuclear spin impurities dominate the coherence decays; hence we consider only
decoherence in isotopically purified samples, such as were investigated in \cite{Wolfowicz2013}.
In that case, donor-donor dipolar interactions dominate the decoherence.
For all sorts of quantum baths, cluster methods have proved extremely
successful in accurately simulating experimental coherence decays 
due to quantum baths \cite{DeSousa,DeSousa2,Witzel1,Maze2008,Renbao2006,Zhao2012,Witzel2013}
for both dipolar coupling and Fermi contact interactions. In this
case we formally consider one of the spins in our pair to be the
spin of interest (``spin A'') and all other non-trivially interacting spins
to be the bath spins (``spin B'') in the usual terminology,
see eg \cite{Raitsimring}. We thus have a number of pairs (all including
spin A). The coherence decay of spin A, $\mathcal{L}(t)= \langle S^+_A \rangle$ 
is thus constructed from the product of all pair contributions: 
\begin{equation}
\mathcal{L}(t) = \prod_{k} {\mathcal{L}}_{k}(t).
\label{clusters}
\end{equation}
For the case where both spins are resonant,
\begin{equation}
 {\mathcal{L}}_{k}(t) = \cos{\frac{J_kt}{4}[(P_u-P_l)^2 \pm 2\rho]}
\label{IDDFF}
\end{equation}
where the $+/-$ corresponds to both spins initially in the same/different
state.

For the case where one spin is not resonant, then
\begin{equation}
 {\mathcal{L}}_{k}(t) = \cos{\frac{J_kt}{2}\rho}
\label{DFF}
\end{equation}

 Eq.\ref{IDDFF} in fact represents the interference of two extremely
well-known spin decoherence mechanisms:\\
 (i) Instantaneous diffusion (ID) is a deleterious effect resulting from the fact that the Hahn pulse also rotates neigbouring spins; thus the effective magnetic noise felt by the central spin changes in time and cannot be refocussed \cite{Raitsimring}.
 The $\frac{J_kt}{4}[(P_u-P_l)^2$ term arising from the diagonal part
of the dipolar interaction gives rise to ID. 
(ii) Direct Flip Flops (DFF) are the result of the off-diagonal 
part of the dipolar coupling and in Eq.\ref{IDDFF} arise from 
the $\frac{J_kt}{2} \rho$ term.\\
These two processes are in some sense two sides of the same coin (the dipolar 
coupling) but are usually considered independently. One key reason is that
in ensemble studies with low donor densities, the spins are 100-200 nm apart
\cite{Tyryshkin2012} thus the coupling is weak relative to local Overhauser
fluctuations i.e. $J \ll \Delta$ where $\Delta$ is the energy detuning between 
these spins. This means that the DFF are strongly suppressed (sometimes deliberately,
by applying a field gradient \cite{Tyryshkin2012}), while the ID processes are not. In this case,
\begin{equation}
 {\mathcal{L}}_{k}(t) \simeq \cos{\frac{J_kt}{4}[(P_u-P_l)^2]}
\label{ID}
\end{equation}
Thus at OWPs, where $P_u \simeq P_l$, then ${\mathcal{L}}_{k}(t) \to 1$
and the ID is suppressed quadratically \cite{Wolfowicz2013} as was seen experimentally.
Thus, to see the coherent interference between ID/DFF type processes, the donor
density has to be increased so $J \gtrsim \Delta$. As the observed DFF suppression
is of order of 10, this would only involved increasing the donor spin density by a factor of 
$\sim 10-100$  (at the cost of much shorter coherence times, though).

In addition, the ID/DFF interference effects apparent in Eq.\ref{IDDFF} are
diluted by non-resonant direct flip-flops Eq.\ref{DFF}.  Unfortunately, in the mixing regime,
there are a range of transitions which are forbidden at high $B_0$ which
lead to non-zero coupling via $S_1^+S_2^-$ terms at low fields due to the mixing.
For example , for the $14 \to 7$ transition of spin A, there is some non-zero
probability of DFF at low fields if the neighbouring
spin B is in any of the states $i=5,6,8,12,13,15$. 
 Hence exposing these interference effects will require a careful numerical calculations.

{\em Conclusions:} We investigate the suppression of dipolar coupling, 
between two electronic spin systems of certain donors in silicon, at '``magic'' 
magnetic field values, and to note possible applications for quantum information. The
``tunable'' nature of the dipolar couplings  as a function of magnetic field 
also suggests the possibility that other field values may offer 
advantage for manipulating larger groups of spins.

Acknowledgements: the author is grateful to S. Bose, S. Balian, G Wolfowicz and R.Guichard for helpful discussions.


\begin{thebibliography}{99}

\bibitem{Nielsen} Quantum Computation and Quantum Information, 
M.A Nielsen, I.L. Chuang,  Cambridge (2000).
\bibitem{Schweiger}Principles of pulse electron paramagnetic resonance spectroscopy,
Schweiger, A. and Jeschke, G. Oxford University Press, Oxford (2001).

\bibitem{Hahn} E. L. Hahn, Phys. Rev. 80, 580 (1950).

\bibitem{Rhim1970} J.S.Waugh, L.M.Huber and U.Haeberlen,Phys. Rev. Lett.,20, 180 (1968);
W-K. Rhim, A. Pines, and J. S. Waugh Phys.Rev. B \textbf{3} 684 (1970).

\bibitem{NV1} L. Robledo, L. Childress, H. Bernien, B. Hensen, P. F. A. Alkemade and R. Hanson, Nature \textbf{477} 574–578 (2011).


\bibitem{NV2} N. Zhao, J. Honert, B. Schmid, M. Klas, J. Isoya, M. Markham, D. Twitchen, F. Jelezko, R.-B. Liu, H. Fedder and J. Wrachtrup, Nature Nano. \textbf{7} 657–662 (2012).
\bibitem{NV2a} P. Cappellaro, L. Jiang, J. S. Hodges and M. D. Lukin, Phys. Rev. Lett.  \textbf{102} 210502 (2009). 

\bibitem{NV3} S. Kolkowitz, Q. P. Unterreithmeier, S. D. Bennett and  M. D. Lukin, Phys. Rev. Lett. \textbf{109} 137601 (2012). 


\bibitem{NV4}  H. Bernien, B. Hensen, W. Pfaff,	G. Koolstra, M. S. Blok, L. Robledo, T. H. Taminiau, M. Markham, D. J. Twitchen, L. Childress and R. Hanson, Nature \textbf{497} 86–90 (2013).
 
\bibitem{NV5} N. Bar-Gill, L. M. Pham, A. Jarmola, D. Budker and R. L. Walsworth, Nature Commun. \textbf{4} 1743 (2013). 
\bibitem{Taminiau} T. H. Taminiau, J. Cramer, T. van der Sar, V. V. Dobrovitski and R. Hanson, Nature Nano. {\bf 9} 171–176 (2014).


\bibitem{Morello2010}  A. Morello, J. J. Pla, F. A. Zwanenburg, K. W. Chan, K. Y. Tan, H. Huebl, M. M\"ott\"onen, C. D. Nugroho, Changyi Yang, J. A. van Donkelaar, A. D. C. Alves, D. N. Jamieson, C. C. Escott, L. C. L. Hollenberg, R. G. Clark and A. S. Dzurak Nature {\bf 467} 687 (2010).

\bibitem{Pla2012} J.J. Pla, K.Y. Tan, J.P. Dehollain, W-H. Lim, J.J.L. Morton, D.N. Jamieson, A.S. Dzurak and A. Morello, Nature 489, 541 (2012). 

\bibitem{Pla2014} J. J. Pla, F. A. Mohiyaddin, K. Y. Tan, J P Dehollain, R. Rahman,
G. Klimeck, D. N. Jamieson, A. S. Dzurak and A. Morello, arxiv:1408.1347 (2014).



\bibitem{Neumann} P. Neumann , R. Kolesov  B. Naydenov, J. Beck, F. Rempp, M. Steiner, V. Jacques, G. Balasubramanian, M. L. Markham, D. J. Twitchen, S. Pezzagna3, J. Meijer , J. Twamley , F. Jelezko  \& J. Wrachtrup, Nature Physics \textbf{6}, 249  (2010).


\bibitem{Tyryshkin2012} A.~M.~Tyryshkin, S.~Tojo, J.~J.~L.~Morton, H.~Riemann, N.~V.~Abrosimov, P.~Becker, H.-J.~Pohl, T.~Schenkel, M.~L.~W.~Thewalt, K.~M.~Itoh and S.~A.~Lyon, Nature Mater. {\bf 11} 143 (2012).


\bibitem{Steger2012} M.~Steger, K.~Saeedi, M.~L.~W.~Thewalt, J.~J.~L.~Morton, H.~Riemann, N.~V.~Abrosimov, P.~Becker and H.-J.~Pohl, Science {\bf 336} 1280 (2012).

\bibitem{Morley2010} G W Morley, M Warner, A M Stoneham, P T Greenland, J van Tol, C W M Kay \& G Aeppli,
Nature Materials 9, 725–729 (2010).

\bibitem{George2010} R. E. George, W. Witzel, H. Riemann, N. V. Abrosimov, N. Notzel, M. L. W. Thewalt and J. J. L. Morton, Phys. Rev. Lett. {\bf 105} 067601 (2010).


\bibitem{Mohammady2010} M.~H.~Mohammady, G.~W.~Morley and T.~S.~Monteiro, Phys. Rev. Lett. {\bf 105} 067602 (2010).

\bibitem{Mohammady2012} M. H. Mohammady, G. W. Morley, A. Nazir and T. S. Monteiro, Phys. Rev. B {\bf 85} 094404 (2012).

\bibitem{Balian2012} S.~J.~Balian, M.~B.~A.~Kunze, M.~H.~Mohammady, G.~W.~Morley, W.~M.~Witzel, C.~W.~M.~Kay and T.~S.~Monteiro, Phys. Rev. B {\bf 86} 104428 (2012).






\bibitem{Wolfowicz2013} G.~Wolfowicz, A.~M.~Tyryshkin, R.~E.~George, H.~Riemann, N.~V.~Abrosimov, P.~Becker, H.-J.~Pohl, M.~L.~W.~Thewalt, S.~A.~Lyon and J.~J.~L.~Morton, Nature Nano. \textbf{8} 561 (2013).

\bibitem{Balian2014} S. J. Balian, G. Wolfowicz, J. J. L. Morton and T. S. Monteiro, Phys. Rev. B \textbf{89} 045403 (2014).

\bibitem{dimers}  J.P. Dehollain, J.T. Muhonen, K.Y. Tan, A. Saraiva, D.N. Jamieson, A.S. Dzurak and A. Morello, Physical Review Letters 112, 236801 (2014); S. Shankar, A. M. Tyryshkin, S. A. Lyon arXiv:1409.3534 (2014).

\bibitem{Thewalt2010} T Sekiguchi, M Steger, K Saeedi, MLW Thewalt, H Riemann, NV Abrosimov, N Nötzel, 
Phys. Rev. Lett. 104, 137402  (2010).

\bibitem{DeSousa} R. de Sousa and S. Das Sarma, Phys. Rev. B \textbf{67} 033301 (2003)

\bibitem{DeSousa2} R. de Sousa and S. Das Sarma,  Phys. Rev. B \textbf{68} 115322 (2003).
 
\bibitem{Witzel1} W. M. Witzel and S. Das Sarma, Phys. Rev. B \textbf{77} 165319 (2008).

\bibitem{Maze2008} J.~R.~Maze, J.~M.~Taylor, and M.~D.~Lukin, Phys. Rev. B {\bf 78}, 094303 (2008).


\bibitem{Renbao2006} R.-B. Liu, W. Yao and L. J. Sham, New J. Phys. \textbf{9} 226 (2007); W. Yao, R.-B. Liu, and L. J. Sham, Phys. Rev. Lett. \textbf{98} 077602 (2007).

\bibitem{Zhao2012} N.~Zhao, S.~W.~Ho, and R.~B.~Liu, Phys. Rev. B {\bf 85}, 115303 (2012).

\bibitem{Witzel2013} W. M. Witzel, M. S. Carroll, \L. Cywi\'nski and S. Das Sarma, Phys. Rev. B \textbf{86} 035452 (2012).

\bibitem{Raitsimring} K. M. Salikhov, S. A. Dzuba, and A. M. Raitsimring,
Journal of Magnetic Resonance, 42, 255-276 (1981).
\end{thebibliography}
\end{document}